\documentclass[prl,twocolumn
			,superscriptaddress ]{revtex4-1}
\usepackage{graphicx}
\usepackage{url}
\usepackage{amsmath}
\usepackage{amssymb}
\usepackage{bbold}
\usepackage{xr-hyper}
\usepackage{hyperref}
\usepackage{bm}
\usepackage{dcolumn}

\usepackage{xcolor}

\newcommand{\rd}{\mathrm{d}}

\newcommand{\Ho}{H}
\newcommand{\Uo}{U}

\newcommand{\be}{\begin{equation}}
\newcommand{\ee}{\end{equation}}
\newcommand{\bes}{\begin{eqnarray}}
\newcommand{\ees}{\end{eqnarray}}
\newcommand{\ket}[1]{{\left|  #1 \right\rangle}}
\newcommand{\bra}[1]{{\left\langle  #1 \right|}}

\begin{document}

\title{Is there a Floquet Lindbladian?}

\date{\today}

\author{Alexander Schnell}
\email{schnell@pks.mpg.de}
\affiliation{Max-Planck-Institut f\"ur Physik komplexer Systeme, 
N\"othnitzer Str.\ 38, 01187 Dresden, Germany}         
\affiliation{Center for Theoretical Physics of Complex Systems,
IBS, Daejeon 305-732, Korea}

\author{Andr\'e Eckardt}
\email{eckardt@pks.mpg.de}
\affiliation{Max-Planck-Institut f\"ur Physik komplexer Systeme, 
N\"othnitzer Str.\ 38, 01187 Dresden, Germany}         
\affiliation{Center for Theoretical Physics of Complex Systems,
IBS, Daejeon 305-732, Korea}

\author{Sergey Denisov}
\email{sergiyde@oslomet.no}
\affiliation{Department of Computer Science, OsloMet--Oslo Metropolitan University, 0130 Oslo, Norway}         
\affiliation{Center for Theoretical Physics of Complex Systems,
IBS, Daejeon 305-732, Korea}
%\affiliation{Lobachevsky State University, Gagarina Av. 23, Nizhny Novgorod, 603950, Russia}

\begin{abstract}
%123456789%123456789%123456789%123456789%123456789%123456789%123456789%123456789%123456789%123456789
The stroboscopic evolution of a time-periodically driven isolated quantum system
can always be described by an effective time-independent Hamiltonian. 
Whether this concept can be generalized to open Floquet
systems, described by a Markovian master equation with time-periodic Lindbladian
generator, remains an open question. By using a two level system as a model, we  explicitly show the existence
of two well-defined parameter regions.
In one region the stroboscopic evolution can be described by a Markovian master
equation with a time-independent Floquet Lindbladian. In the other  it cannot;
but here the one-cycle evolution operator can be reproduced with an effective non-Markovian master equation
that is homogeneous but non-local in time. Interestingly, we find that the
boundary between the phases depends on when  the evolution is stroboscopically monitored. 
This reveals the non-trivial role played by the micromotion in the dynamics of 
open Floquet systems.  
\end{abstract}
%\pacs{}
%\keywords{}

\maketitle
When the coherent evolution of an isolated quantum Floquet system, described by the time-periodic 
Hamiltonian $H(t)=H(t+T)$, is monitored stroboscopically in steps of the driving period $T$, this 
dynamics is described by repeatedly applying the one-cycle time-evolution operator 
$U(T) = \mathcal{T} \exp\big[-\frac{i}{\hbar}\int_{0}^T\!\rd t'H(t')\big]$ (with time ordering
$\mathcal{T}$) \cite{Shirley65,Sambe73}. It can always be expressed in terms of an effective
time-independent Hamiltonian $H_F$, called Floquet Hamiltonian, $U(T) \equiv \exp(-i H_F T/\hbar)$. 
While the Floquet Hamiltonian is not unique due to the multi-branch structure of the operator logarithm
$\log U(T)$, the unitarity of $U(T)$ implies that $H_F$ is Hermitian (like a proper Hamiltonian) for every branch. The concept
of the Floquet Hamiltonian suggests a form of quantum engineering, where a suitable time-periodic driving
protocol is designed in order to effectively realize a system described by a Floquet Hamiltonian with 
desired novel properties. This type of Floquet engineering was successfully employed with ultracold 
atoms~\cite{Eckardt17},  e.g.\ to realize artificial magnetic fields and topological band structures
for charge neutral particles \cite{AidelsburgerEtAl11, RechtsmanEtAl13, StruckEtAl13, JotzuEtAl14, 
AidelsburgerEtAl15, Flaeschner16}. 
%[Aidelsburger 11, Rechtsman13, Struck13, Jotzu14, Aidelsburger14, Fläschner 16.]. 

However, systems like atomic quantum gases, which are very well isolated from their environment, should rather be 
viewed as an exception. Many quantum systems that are currently studied in the laboratory and used for
technological applications are based on electronic or photonic degrees of freedom that usually couple to their environment.
It is, therefore, desirable to extend the concept of Floquet engineering also to open systems. 
In this context, a number of papers investigating properties of the non-equilibrium steady states approached by 
periodically modulated dissipative systems in the long-time limit have been published  \cite{BreuerEtAl00, AlickiEtAl06, KetzmerickWustmann10, VorbergEtAl13,
ShiraiEtAl14, SeetharamEtAl15, SeetharamEtAl15, DehghaniEtAl15, IadecolaEtAl15,VorbergEtAl15, ShiraiEtAl16, 
LetscherEtAl17, SchnellEtAl18, chong2018, QinEtAl18, HigashikawaEt18}. 
% i.e.\ using Floquet-Born-Markov theory \cite{BluemelEtAl91, KohlerEtAl97, HoneEtAl09}. 
%[Sergey-Juzar-Paper, 
%our Work, Papers on open Floquet topological insulators, \ldots]. 
In this paper, in turn,  we are interested in the (transient) dynamics of open Floquet systems 
%including the transient evolution on the way to the asymptotic state)
and address the question as to whether it is possible to describe their stroboscopic evolution 
%in  terms of an effective 
with time-independent generators that generalize the concept of the Floquet Hamiltonian to open systems. 

We consider a time-dependent Markovian master equation \cite{BreuerEtAl09}
\be\label{eq:tdm}
\dot{\rho} = \mathcal{L}(t)\rho = \frac{1}{i\hbar}[H(t),\rho] + \mathcal{D}(t)\rho,
\ee
for the system's density operator $\rho$ (with Hilbert space dimension $N$), described by a time-periodic generator
$\mathcal{L}(t)=\mathcal{L}(t+T)$. It is characterized by a Hermitian time-periodic
Hamiltonian $H(t)$ and a dissipator 
\be
\mathcal{D}(t)\rho = 
\sum_i  \big[A_i(t)\rho A_i^\dag(t)-\frac{1}{2}\{A_i^\dag(t)A_i(t),\rho\}\big],
\ee
with traceless time-periodic jump operators $A_i(t)$. The generator $\mathcal{L}$ is of
Lindblad form \cite{Lindblad1976} (it is a Lindbladian). This is the most general time-local form guaranteeing a completely positive and trace preserving (CPTP) map consistent
with quantum mechanics that is (time-dependent) Markovian \cite{BreuerEtAl09,HallEtAl14} (in the sense of that it is CP-divisibile).
 %\footnote{One might call this time-dependent Markovian. Note that there }). 
In particular, the one-cycle evolution superoperator 
\be\label{eq:oceo}
\mathcal{P}(T)=\mathcal{T}\exp\bigg[\int_0^T\!\rd t\, \mathcal{L}(t) \bigg],
\ee
the repeated application of which describes the stroboscopic evolution of the system, is CPTP.
%Now the concrete question to be addressed is, whether we can again, like for isolated Floquet systems, 
%interpret this stroboscopic evolution as the evolution of an effective time-independent master equation. 

\begin{figure}%Markovian
	\includegraphics[width=0.99\columnwidth]{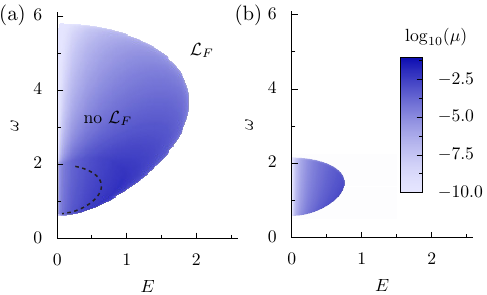}
	\caption{Distance from Markovianity $\mu$ of the effective generator of the one-cycle evolution superoperator 
as a function of driving strength $E$ and  frequency $\omega$, for weak dissipation $\gamma = 0.01$ and two
driving phases (a) $\varphi = 0$  and (b) $\varphi = \pi/2$. In the white
region, where $\mu=0$, a Floquet Lindbladian $\mathcal{L}_F$ exists. 
On the dashed line the Floquet map $\mathcal{P}(T)$ possesses two negative real eigenvalues.}
	\label{fig:main}
\end{figure}

We can now distinguish three different possible scenarios for a given time-periodic Lindbladian
$\mathcal{L}(t)$: 
(a) the action of $\mathcal{P}(T)$ can be reproduced with an effective (time-independent) Markovian master equation described by 
a time-independent generator of Lindblad form (Floquet Lindbladian)~$\mathcal{L}_F$, $\mathcal{P}(T)=\exp(T\mathcal{L}_F)$; 
(b) the action of $\mathcal{P}(T)$ is reproduced with an effective non-Markovian master equation characterized by a 
time-homogeneous memory kernel;
(c) neither (a) nor (b), i.e.~the action of 
$\mathcal{P}(T)$ cannot be reproduced with any time-homogeneous master equation. 
Scenario (a) is implicitly assumed in recent papers \cite{HaddadfarshiEtAl15,RestrepoEtAl17,DaiEtAl16}, 
where a high-frequency Floquet-Magnus expansion \cite{BlanesEtAl09} (routinely used for isolated 
Floquet systems \cite{GoldmanDalibard14,BukovEtAl15,EckardtAnisimovas15}) is employed in order to construct 
an approximate Floquet Lindbladian. It requires that at least one branch of the operator logarithm 
$\log \mathcal{P}(T)$ has to be of Lindblad form so that it can be associated with $T\mathcal{L}_F$. 
However, differently from the case of isolated systems, 
%for which any logarithm branch of the unitary evolution 
%operator is guaranteed to be Hermitian, 
it is not obvious whether there is at least one valid branch for a given 
open Floquet system, since general CPTP maps do not always possess a logarithm of Lindblad type~\cite{WolfEtAl08}.
Below we demonstrate that scenario (a) is not always realized even in the case of a simple two-level model.
Instead, we find that the parameter space is shared by two phases corresponding to scenario (a) and (b), respectively.

%fulfilled by using a simple 
%In the following, we illustrate that scenario (a) is indeed not always fulfilled by using a simple 
%model, for which we find two phases corresponding to scenario (a) and (b), respectively.   
%In the latter phase we explicitly construct an effective time-homogeneous memory kernel and compute 
%different measures of non-markovianity for the non-Lindbladian generator $\log[\mathcal{P}(T)]$ (picking 
%the branch of the logarithm closest to markovianity).

We consider a two-level system described by a time-periodic Hamiltonian $H(t)$ and a single 
time-independent jump operator $A$, 
%which read
\begin{align}\label{eq:ham}
%H(t) = \Delta \sigma_z + E\cos(\omega t)\sigma_x,
H(t) = \frac{\Delta}{2} \sigma_z + E\cos(\omega t-\varphi)\sigma_x
\text{  and  } A=\sqrt{\gamma}\sigma_-.
\end{align}
Here $\sigma_x$, $\sigma_z$ and $\sigma_-$ are standard Pauli and lowering operators. Using
the level splitting $\Delta$ and $\hbar/\Delta$ as units for energy and time (so that henceforth 
$\Delta=\hbar=1$), the model is characterized by four dimensionless real parameters: 
the dissipation strength  $\gamma$ as well as the driving strength $E$, frequency $\omega$, and phase $\varphi$. 
%the driving strength $E$, frequency $\omega$, and phase $\varphi$, as well as the dissipation strength $\gamma$.

\begin{figure}
	\includegraphics[angle=0,width=0.8\columnwidth]{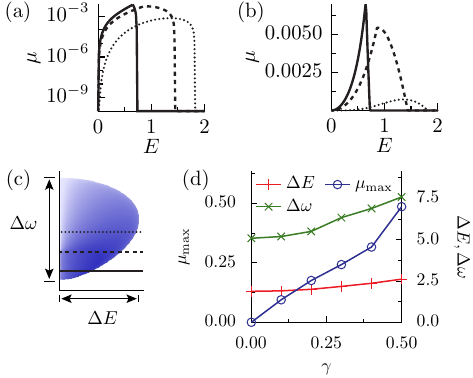}
	\caption{(a, b)~Distance from Markovianity $\mu$ along horizontal cuts 
	through the phase diagram of Fig.~\ref{fig:main}(a) for $\omega=1, 2, 3$ (solid, dashed, dotted line,
	cuts are shown in~(c)). %, in  (a) logarithmic and (b) linear plots.
	(d)~Maximum extent of the non-Lindbladian phase with respect to frequency, $\Delta\omega$, and driving strength, $\Delta E$, (defined as we sketch in~(c))  and maximum non-Markovianity 
	$\mu_\text{max}=\max_{\omega,E}[\mu(\omega,E)]$ versus dissipation strength $\gamma$.}
	\label{fig:gamma-limit}
\end{figure}

%In the coherent limit, a periodically modulated $N$-dimensional quantum system is  typically modeled  with a
%time-periodic Hamiltonian, $H(t+T)=H(t)$. This is a basic set-up of Floquet  physics \cite{Shirley65,Sambe73,GrifoniHaenggi98,GoldmanDalibard14,BukovEtAl15,Eckardt17}, in which  
%a major role is
%played by a Floquet operator $U(T) = \mathcal{T}
%\exp\big[-\frac{i}{\hbar}\int_{0}^T\!\rd t'H(t')\big]$ (here $\mathcal{T}$ is the standard time ordering). 
Let us first address the question of the existence of a Floquet Lindbladian.
For an open system, $\gamma>0$, we have to consider the one-cycle evolution superoperator $\mathcal{P}(T)$ 
\cite{szczygielski2014,HartmannEtAl17}. Since it is a CPTP map, its spectrum is invariant under complex  
conjugation. Thus, its $N^2$ eigenvalues are either real or appear as complex conjugated pairs (we denote 
the number of these pairs $n_c$). { This \textit{Floquet map} %can be represented
%in its Jordan normal form, 
shall be diagonalized,
$\mathcal{P}(T) =    \sum_a \lambda_a \mathcal{M}_{a} =   \sum_r \lambda_r \mathcal{M}_r + \sum_c
(\lambda_c \mathcal{M}_c + {\lambda}^*_c \mathcal{{M}}_{c*})$, with $n_c$ pairs $\{\lambda_c, {\lambda}^*_c \}$ and (not necessarily self-adjoint) projectors $\mathcal{M}_{a}$.

To find out whether we are in scenario (a), we implement the Markovianity test proposed by Wolf \emph{et al.}\
in Refs.~\cite{WolfEtAl08,Cubitt2012}. Namely, in order to be consistent with a time independent Markovian evolution, $\mathcal{P}(T) $ should have at least one logarithm branch, 
%a super-operator 
$\mathcal{S}_{\{\mathbf{x}\}} = \frac{1}{T}\log \mathcal{P}(T) = \mathcal{S}_0 + \frac{2\pi i}{T}\sum_{c=1}^{n_c} x_c (\mathcal{M}_c -  \mathcal{{M}}_{c*})$, that gives rise to a 
valid Lindblad generator
($\mathcal{S}_0$ is the principal branch). Here a set of $n_c$ integers $\{\mathbf{x}\} = \{x_1,...,x_{n_c}\} \in \mathbb{Z}^{n_c}$ labels a branch of the logarithm. 
To get the Floquet Lindbladian~$\mathcal{L}_F$, we should find a branch for which the superoperator
$\mathcal{S}_{\{\mathbf{x}\}}$  fulfills two conditions: (i) it should preserve Hermiticity
%i.e.,
%$\mathcal{S}_{\{\mathbf{x}\}} V^\dagger = \mathcal{S}_{\{\mathbf{x}\}} V$,
%if $V = V^\dagger$, 
and (ii) it has to be \textit{conditionally} completely positive \cite{Evans77}. 
%Then the corresponding branch
%$\mathcal{S}_{\{\mathbf{x}\}}$ can be nominated for an effective Lindblad generator $\mathcal{L}_F$. 
Already here the contrast with the unitary case (where all branches provide a licit Floquet Hamiltonian)
 becomes apparent: it is not guaranteed that such branch exists. 
There is no need to inspect the different branches to check condition (i). It simply demands that the spectrum of
$\mathcal{S}_{\{\mathbf{x}\}}$ has to be invariant under complex conjugation. This means, in turn, that the 
spectrum of the Floquet map $\mathcal{P}(T)$ should not contain negative real eigenvalues (or, strictly speaking, there must be no eigenvalues of odd degeneracy,
whose integer powers give negative numbers). 
%Otherwise, the answer is `no'. If the answer is `yes', we need to check the fulfillment of the condition (ii). 
Condition~(ii) is more complicated and involves properties of the spectral projectors $\mathcal{M}_a$ %eigenelements 
of the Floquet map. 
The corresponding test was formulated in Refs.~\cite{WolfEtAl08,Cubitt2012} (we provide a brief operational description in the supplemental material~\cite{SM})}.

If the result of one of the two tests is negative and therefore no Floquet Lindbladian exists, it is instructive to quantify the distance 
from Markovianity by introducing some measure and then picking the branch $\mathcal{S}_{\{\mathbf{x}\}}$ giving rise to the minimal value of the measure. 
For this purpose, we compute two different measures for non-Markovianity 
proposed by Wolf \emph{et al.}~\cite{WolfEtAl08} and Rivas \emph{et al.}~\cite{RivasEtAl10}. 
The first measure is based on adding a noise term $-\chi\mathcal{N}$ of strength $\chi$ to 
the generator and noting the minimal strength $\mu = \chi_\text{min}$ required to make it Lindbladian [so that it fulfills conditions (i) and (ii)]. Here
$\mathcal{N}$ is the generator of the depolarizing map
$\exp(T\chi \mathcal{N})\rho = \mathrm{e}^{-\chi T} \rho + [1 - \mathrm{e}^{-\chi T}] \frac{\mathbb{1}}{N}$ 
\cite{WolfEtAl08, SM}. 
%In words, the evolution is diluted with noise, by continuously increasing the noise strength. Once the evolution becomes Markovian, the corresponding noise  strength is used as the measure of the distance to Markovianity.
The second measure quantifies the violation of positivity of the Choi representation
\cite{Choi75,Jamiolkowski72,Holevo2012} of the generated map \cite{SM,RivasEtAl10}. Interestingly, we find 
that for our model system both measures agree: within the numerical accuracy the second measure is always found 
to be equal to $\mu/2$.

%Let us now discuss our results. 
In Fig.~\ref{fig:main}(a) we plot the distance from Markovianity
$\mu$ for the effective generator of the one-cycle evolution superoperator $\mathcal{P}(T)$ versus driving 
amplitude $E$ and frequency $\omega$. We choose $\varphi =0$ and weak dissipation $\gamma=0.01$. The 
%extended 
blue lobe, where $\mu > 0$, corresponds to
%an extended 
a phase, where a Floquet Lindbladian $\mathcal{L}_F$ does not exist. This non-Lindbladian phase is  surrounded by a Lindbladian 
phase (white region) where $\mu = 0$ so that $\mathcal{L}_F$ can be constructed [scenario (a)].
It contains also the $\omega$ axis, corresponding to the trivial undriven limit $E=0$. Note that only 
for a fine-tuned set of parameters, lying on the dashed line in Fig.~\ref{fig:main}(a), 
$\mathcal{P}(T)$ possesses negative eigenvalues. However, they come in a degenerate pair, such that
the 
construction of a Floquet Lindbladian is not hindered by condition (i). Both the high- and the low-frequency limit are 
surrounded by finite frequency intervals, where the Floquet Lindbladian exists. 
This suggests that it might be possible to construct the Floquet Lindbladian in the high-frequency
regime from a Floquet-Magnus-type expansion \cite{HaddadfarshiEtAl15,RestrepoEtAl17,
DaiEtAl16}. 
%This suggests that a 
%Floquet-Magnus-type expansion for the Floquet-Lindbladian \cite{HaddadfarshiEtAl15,RestrepoEtAl17,
%DaiEtAl16} can describe the high-frequency regime. 
Somewhat counter-intuitively, we find that 
the Floquet Lindbladian always exists for sufficiently strong driving strengths $E$, so that for large $E$ the low and the high-frequency Lindbladian phases are connected. 
However, for intermediate frequencies, a phase where no Floquet Lindbladian exists stretches over a finite interval of 
driving strengths $E$ separated only infinitesimally from the undriven limit $E=0$. This can also be seen from 
Fig.~\ref{fig:gamma-limit}(a) and (b), where we plot $\mu$ along horizontal cuts through the
phase diagram [indicated by the lines of unequal style in Fig.~\ref{fig:gamma-limit}(c)] using a logarithmic and a linear scale, respectively.  

Figure~\ref{fig:main}(b) shows the phase diagram for a different driving phase, $\varphi=\pi/2$. Remarkably, 
compared to $\varphi=0$ [Fig.~\ref{fig:main}(a)] the non-Lindbladian phase now covers a much smaller area 
in parameter space. The phase boundaries depend on the driving phase or, in other words, on when during the 
driving period we monitor the stroboscopic evolution of the system in a particular experiment. In the coherent limit, we can decompose 
the time evolution operator of a Floquet system from time $t_0$ to time $t$ like 
$\Uo(t,t_0)=U_F(t)\exp[-i(t-t_0)H_\text{eff}]U_F^\dag(t_0)$, where $\Uo_F(t)=\Uo_F(t+T)$ is a
unitary operator describing the time-periodic micromotion of the Floquet states of the system and 
$H_\text{eff}$ is a time-independent effective Hamiltonian. The Floquet Hamiltonian $\Ho^F_{t_0}$, defined 
via $U(t_0+T,t_0)=\exp(-i T\Ho^F_{t_0})$ so that it describes the stroboscopic evolution of the system at 
times $t_0$, $t_0+T$, \ldots, is for general $t_0$ then given by $H^F_{t_0}
=\Uo_F(t_0)\Ho_\text{eff}\Uo_F^\dag(t_0)$ \cite{EckardtAnisimovas15}. (Note that above we used the lighter notation
$\Ho_F=\Ho^F_0$ for $t_0=0$.) It depends on the micromotion via a $t_0$-dependent unitary rotation. However, in the dissipative system the micromotion will no longer be captured by a unitary operator.
This explains why the effective time-independent generator of the stroboscopic evolution can change its
character as a function of $t_0$ (or, equivalently, the driving phase $\varphi$) in a nontrivial fashion, e.g.\ from Lindbladian to non-Lindbladian. 
%Note that this $t_0$-dependence of the non-Lindbladian region is not ambiguous because it corresponds to different experimental protocols.

In Fig.~\ref{fig:gamma-limit}(d), the dependence of the phase diagram 
on the dissipation strength $\gamma$ is investigated. % in more detail. 
We find that the extent of the non-Lindbladian phase both in 
frequency, $\Delta\omega$, and driving strength, $\Delta E$, [defined in Fig.~\ref{fig:gamma-limit}(c)] 
does not vanish in the limit $\gamma\to0$.
Thus, even for arbitrary weak dissipation the Floquet Lindbladian does not exist in a substantial region of 
parameter space. It is noteworthy that the maximum distance from Markovianity $\mu$ goes to zero linearly with $\gamma$, i.e., 
the non-Markovianity is a first-order effect with respect to the dissipation strength.

While in the non-Lindbladian phase, we are not able to find a Markovian time-homogeneous master equation 
reproducing the one-cycle evolution operator $\mathcal{P}(T)$, one might still be able to construct a 
time-homogeneous non-Markovian master equation, which is non-local in time and described by a memory kernel \cite{Budini04,BreuerEtAl16RMP,Chruscinski16}. In  order to construct such an equation, we assume an evolution 
with an exponential memory kernel for $t\leq T$
\begin{equation}
	\partial_t  \tilde{\varrho}(t) = \frac{1}{\tau_\text{mem}} \int_{0}^t d \tau \, e^{(\tau-t)/\tau_\text{mem}} 
		\mathcal{K}\tilde{\varrho}(\tau),
	\label{eq:eom-kernel}
\end{equation}
where $\tau_\text{mem}$ is the memory time and $\mathcal{K}$ the kernel superoperator. 
%The kernel on the right-hand side is not of arbitrary form. In the Laplace domain it is linked to the 
%probability density 
%function, which governs the time between two successive applications of a CPTP map $\mathcal{E}=\id+\mathcal{L}_K$ to a system performing a single realization of a stochastic microscopic process
%\cite{Budini04}. Thus, the CPTP character of the evolution, averaged over many 
%realizations and resulting in Eq.~\eqref{eq:eom-kernel}, is guaranteed.
It is important to understand  that a time-homogeneous master 
equation~(\ref{eq:eom-kernel}), when being integrated forward in time $t$ also beyond  $T$, would not reproduce the same map
after every period,  since $\tilde{\mathcal{P}}(2T)$, $\tilde{\mathcal{P}}(3T)$, etc., will depend on the 
the corresponding pre-history of the length $2T$, $3T$, etc.. The stroboscopic evolution can only be obtained by erasing the memory 
after every period, which formally corresponds to multiplying the integrand of Eq.~(\ref{eq:eom-kernel}) by 
$\Theta\left(\tau-\lfloor t/T \rfloor T\right)$, where $\Theta$ and $\lfloor \cdot \rfloor $ 
denote the Heaviside step function and the floor function, respectively.

\begin{figure}
	\includegraphics[angle=0,width=0.95\columnwidth]{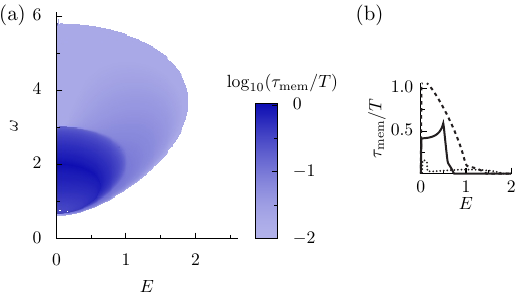}
	\caption{(a)~Shortest memory time $\tau_{\text{mem}}$ for the exponential kernel of the effective
	non-Markovian generator in Eq.~\eqref{eq:eom-kernel}. $\tau_\text{mem}=0$ (white) indicates
	the Lindbladian phase. Due to limited numerical accuracy, we cannot resolve values of $\tau_{\mathrm{mem}} \le 10^{-2}T$. This leads to a spurious plateau of apparently constant $\tau_\text{mem}=10^{-2}T$. 
	Other parameters as in Fig.~\ref{fig:main}(a). (b)~Cuts through the phase diagram at $\omega=1,2,3$ (solid, dashed, dotted line), similar to Fig.~\ref{fig:gamma-limit}(b).}
	\label{fig:t-mem}
\end{figure}

Let the map $\tilde{\mathcal{P}}$ describe the evolution
resulting from the effective master equation~(\ref{eq:eom-kernel}), 
$\tilde\varrho(t) = \tilde{\mathcal{P}}(t) \varrho(0)$. It solves $\partial_t \tilde{\mathcal{P}}(t)= 
{\tau}^{-1}_\text{mem}\int_{0}^t \mathrm d \tau \, e^{(\tau-t)/\tau_\text{mem}}  \mathcal{K}\tilde{\mathcal{P}}(\tau)$ with
$\tilde{\mathcal{P}}(0) = \mathbb{1}$. 
We now have to construct a superoperator $\mathcal{K}$, so that $\tilde{\mathcal{P}}(T)=\mathcal{P}(T)$.
For that purpose, we represent the one-cycle evolution in its diagonal form,
$\mathcal{P}(T) =  \sum_a \lambda_a \mathcal{M}_{a}$. 
%with orthogonal (non-hermitian) projectors $P_{a}$ onto the eigenspace $a$
%with corresponding (complex) frequency $\Omega_a= \Delta\omega_a-i\gamma_a$.
A natural ansatz is then $\mathcal{K} = \sum_a \eta_a \mathcal{M}_{a}$, for which  
we find an evolution operator of the form 
$\tilde{\mathcal{P}}(t) = \sum_a h_{a}(t) \mathcal{M}_{a}$, with characteristic decay functions $h_a(t)$ 
obeying $h_a(0) = 1$. Plugging this ansatz into the equation of motion, the problem reduces to solving a 
set of  scalar equations.
%$\partial_t h_{a}(t) = \int_{0}^t \mathrm d \tau \, \mathrm{e}^{(\tau-t)/\tau_\text{mem}} \lambda^K_a h_{a}(\tau)$. 
They possess solutions \cite{SM}  
$h_{a}(t) = \mathrm{e}^{-t/2\tau_\text{mem}} [ \cosh(\Gamma_a t) + \sinh(\Gamma_a t) /(2\Gamma_a\tau_\text{mem})]$ with $\Gamma_a = [\tau_\text{mem}^{-2}/4 +\tau_\text{mem}^{-1} \eta_a]^{1/2}$.
Requiring $	h_{a}(T) = \lambda_a$, determines the eigenvalues $\eta_a$ as a function of the memory time
$\tau_\text{mem}$. It is then left to check, whether the corresponding $\mathcal{K}=\mathcal{K}(\tau_{\mathrm{mem}})$, which depends on the
memory time $\tau_\text{mem}$, gives rise to an evolution that is CPTP at all times $t$.
Note that in contrast to the Markovian limit, $\tau_\text{mem} \to 0$, where $\mathcal{K}$ needs to be of Lindblad form,
for finite memory time $\tau_\text{mem}$ it is an intriguing open question to find general conditions that
characterize the admissible superoperators $\mathcal{K}$ that give rise to an evolution that is CPTP. 
Ideas to characterize special cases \cite{Budini04,Maniscalco07,SiudzinskaEtAl17,SiudzinskaEtAl19,WudarskiEtAl15}
have been developed but unfortunately are not directly applicable to our problem. Even though general sufficient conditions exist \cite{ChruscinskiKossakowski16}, 
it is unclear how to bring Eq.~\eqref{eq:eom-kernel}
into a form that is required to prove these conditions.

In the absence of a general criterion, we perform a  numerical test to check whether $\tilde{\mathcal{P}}(t)$ is completely 
positive.  The test is based on the fact that  a given map $\mathcal{P}$ 
 is completely positive if and only if its Choi representation is positive, $\mathcal{P}^\Gamma = (\mathcal{P} \otimes \rm id)\ket{\Omega}\bra{\Omega} \geq 0$ 
 \cite{Choi75, Jamiolkowski72}, where $\ket{\Omega} = \frac{1}{\sqrt{2}} \left( \ket{00} + \ket{11} \right)$ is a maximally entangled state of the system  and an ancilla of the same size.
Thus, we require positivity of the Choi representation, 
$\tilde{\mathcal{P}}(t_n)^\Gamma\geq 0$, for all times $t_n \in [0, T]$
on a numerical grid with 100 intermediate steps.
{Note that, because of the memory erasure after every period,  we do not impose the CPTP condition on the maps generated by the pair $\{\mathcal{K}(\tau_\text{mem}), \tau_\text{mem}\}$
for times $t > T$.}

%, is of Lindblad form by performing the test for condition (ii). 
For all parameters, we find a memory time $\tau_\text{mem}$ such that  $\mathcal{K}$
gives rise to an evolution that is CPTP. 
In the phase, where the Floquet Lindbladian $\mathcal{L}_F$ exists, we find a kernel
$\mathcal{K}$ which yields a CPTP evolution for arbitrarily short memory times $\tau_\text{mem}$. 
In contrast, in the non-Lindbladian phase 
the memory time $\tau_\text{mem}$ cannot be smaller than some minimal value. In Fig.~\ref{fig:t-mem}(a) we plot 
this minimal memory time versus driving strength and frequency. The resulting map shows good qualitative 
agreement with the distance to Markovianity $\mu$ shown in Fig.~\ref{fig:main}(a) %(the apparent plateau for small values of $\tau_\text{mem}$ 
(the apparent plateau of constant $\tau_\text{mem}=10^{-2}T$
is an artifact related to the fact 
that our numerical implementation is not able to resolve memory times smaller than $10^{-2} T$). Nevertheless, in contrast to the measure $\mu$,  $\tau_{\text{mem}}$ does not tend to zero for small $E$ 
as we observe in Fig.~\ref{fig:t-mem}(b). %to Fig.~\ref{fig:gamma-limit}(b).  
It is possible that a different behavior of $\tau_\text{mem}$ would be found for a more general ansatz of the memory kernel. 
The specific form of our ansatz implies that the minimal memory time found here 
provides an upper bound for the minimal memory time for general time-homogeneous memory kernels only. 
Note that  % because of the memory erasure after each period, 
the memory time
$\tau_\mathrm{mem}$ can even be larger than $T$.

%Additionally, we report an important observation (see Ref.~\cite{SM} for more details): Even though general CPTP-guaranteeing conditions for
%the shape of $\mathcal{K}$ are unknown in our non-Markovian case, for the model (4)  we observe that 
%the minimal memory time $\tau_\mathrm{mem}$ can be gauged  by requiring that the kernel $\mathcal{K}$ is of Lindblad
%form [this could be checked by performing  the test for condition (ii)\cite{SM}  on $\mathcal{K}$].
%For \emph{large} values of $\tau_\mathrm{mem}$ we observe that it is even enough to require that $\mathcal{K}$ has to be  of 
%Lindblad
%form in order to guarantee that the evolution $\tilde{\mathcal{P}}(t)$ is CPTP. 
Interestingly, we find that in the regime of \emph {large} memory times $\tau_\text{mem}$, the minimal memory time is found for a Kernel operator $\mathcal{K}$ of Lindblad form. 
%At first this seems counter-intuitive because the Markovian limit is for short $\tau_\mathrm{mem}$ 
%\textbf{(not clear. Alex: could you please formulate it better?)}.
At first this seems counter-intuitive because Eq.~\eqref{eq:eom-kernel} gives rise to a Markovian evolution in the opposite limit,
$\tau_\mathrm{mem}\to 0$.
However one can show \cite{SM} $\tilde{\mathcal{P}}(t)\approx  \exp({\mathcal{K}}t^2/{2\tau_\mathrm{mem}})$ for $t \ll \tau_\mathrm{mem}$, 
which is a quantum semigroup with rescaled time $t' = t^2$.
Thus, the map $\tilde{\mathcal{P}}(t)$ is guaranteed to become CPTP in the limit   $t/\tau_\text{mem}\to0$ for a Lindbladian Kernel $\mathcal{K}$ and can be expected to remain CPTP also for a significant fraction of the period where $t/\tau_\text{mem}\ll1$.
%Thus, CPTP is guaranteed for $t \ll \tau_\mathrm{mem}$ if $\mathcal{K}$ is a Lindbladian, which 
%for large $\tau_\mathrm{mem}$ describes a significant fraction of the period $T$.

%Note that our ansatz for the Kernel superoperator $\mathcal{K}$ in Eq.~\eqref{eq:eom-kernel} is more restrictive
%than one would actually need it. But if we would relax it
%the problem becomes hard to solve even numerically, because
%the propagator $\tilde{\mathcal{P}(T)}$ cannot be calculated so easily as before.
%Therefore, the memory time $\tau_\text{mem}$ that we present here is in principle only an
%upper bound on the shortest memory time for all possible Kernel superoperators $\mathcal{K}$.

Finally, the fact that for the used model, we can always construct a time-homogeneous memory kernel which yields a CPTP evolution on the time interval $[0,T]$, means that the non-Lindbladian 
phase in this case corresponds to scenario (b). 
Nevertheless, let us stress that the stroboscopic action of $\mathcal{P}(nT)$ over more than one period (i.e.\ not only for $n=1$, but
also for all $n\ge2$) can in general 
not be obtained from an effective time-homogeneous non-Markovian evolution like in Eq.~\eqref{eq:eom-kernel}, but with $t$ taking also values $t>T$, 
because it is essential that the memory is erased at stroboscopic instances of time.

Our results shed light on limitations and opportunities for Floquet engineering in open quantum systems. Using
a simple model system, we have shown that an effective Floquet Lindbladian generator, constructed analogously 
to the Floquet Hamiltonian for isolated Floquet systems, exists in extensive parameter regimes. In particular
for sufficiently large driving frequencies the Floquet Lindbladian can be constructed, suggesting that here 
high-frequency approximation schemes \cite{HaddadfarshiEtAl15,RestrepoEtAl17, DaiEtAl16} should indeed be applicable 
(even though it is an open question whether or when these give rise to the correct Lindbladian effective generator). 
However, we found also an extended parameter region, where it does not exist, and where only a time-homogeneous non-Markovian
effective master equation is able to reproduce the one-cycle evolution. 
%This is the most interesting parameter range of strong resonance modulations, very different from the high-frequency
%diabatic limit usually addressed in recent works \cite{HaddadfarshiEtAl15,RestrepoEtAl17, DaiEtAl16}. 
This finding poses an intriguing question as to whether time-dependent Markovian systems can be 
used -- in a controlled fashion -- to mimic non-Markovian %time-homogeneous 
ones. Another relevant observation is
that the existence of the Floquet Lindbladian depends on when during the driving period %-- on the interval $[0,T]$-- 
the model is stroboscopically monitored. This reveals an important role played by the non-unitary micromotion in open 
Floquet systems, which we might hope to exploit for the purpose of dissipative Floquet engineering, and which may as well
be important in the context of quantum heat engines~\cite{ScopaEtAl18}. In future 
work, it will be crucial to develop intuitive approximation schemes allowing to tailor the properties of open
Floquet systems. Also, the behavior of larger systems has to be investigated (though from the computational point
of view it is a very hard problem; see, e.g., Ref.~\cite{HartmannEtAl17} for a first study in this direction). 
%that already condition (i) is not fulfilled in the 
%region of resonant driving because a large fraction of the Floquet eigenvalues occupy the negative part of the 
%real axis.
%The answer to the question we posed as the title is ``sometimes yes, sometimes no''.
%The absence of the effective generator in the most interesting region of resonance non-perturbative driving 
%provokes another question:
%Is the Markovian description of the type Eq.~\eqref{eq:tdm}, with time-periodic Hamiltonian and time-
%independent dissipative Liouvillian, legit? Should the  rates (and even the jump operators)
%be always made time-periodic in order to make  modulated  evolutions `embeddable' into time-independent 
%evolutions? We are not aware of any fundamental principle which 
%enforces this condition. If such principle  does not exist,  modulations serve a principally new type of
%\textit{dissipative} Floquet engineering, allowing to mimic non-Markovian 
%evolution (though with a restricted memory) by using modulated Markovian evolution. 
%In the yes-region another issue is of relevance: What is the  best generalization of the  Magnus technique 
%which guarantees the fastest convergance, especially in terms of 
%the order \cite{ReimerEtAl18}?
%Finally, recent studies of a more complex scalable model with $N \gg 2$ states \cite{HartmannEtAl17},  revealed 
%that already condition (i) is not fulfilled in the 
%region of resonant driving because a large fraction of the Floquet eigenvalues occupy the negative part of the 
%real axis.

\vspace{0cm}

\begin{acknowledgments}
We thank K. \.{Z}yczkowski and D. Chruscinski for fruitful discussions and an anonymous referee
for pointing out a mistake that was present in the first version of the paper. 
S.D.~acknowledges support by  the  Russian  Science Foundation Grant No.~19-72-20086.
A.S.~and A.E.~acknowledge financial support by the DFG via the Research Unit FOR 2414. 
\end{acknowledgments}

\bibliography{mybib}

\end{document}

\title{Supplemental Material for \\ ``Is there a Floquet Lindbladian?''}

\author{Alexander Schnell}
\email{schnell@pks.mpg.de}
\affiliation{Max-Planck-Institut f\"ur Physik komplexer Systeme, 
N\"othnitzer Str.\ 38, 01187 Dresden, Germany}         
\affiliation{Center for Theoretical Physics of Complex Systems,
IBS, Daejeon 305-732, Korea}

\author{Andr\'e Eckardt}
\email{eckardt@pks.mpg.de}
\affiliation{Max-Planck-Institut f\"ur Physik komplexer Systeme, 
N\"othnitzer Str.\ 38, 01187 Dresden, Germany}         
\affiliation{Center for Theoretical Physics of Complex Systems,
IBS, Daejeon 305-732, Korea}

\author{Sergey Denisov}
\email{sergiyde@oslomet.no}
\affiliation{Department of Computer Science, OsloMet--Oslo Metropolitan University, 0130 Oslo, Norway}         
\affiliation{Center for Theoretical Physics of Complex Systems,
IBS, Daejeon 305-732, Korea}
%\affiliation{Lobachevsky State University of Nizhny Novgorod, Gagarina Av. 23, Nizhny Novgorod, 603950, Russia}

%05.10.Gg	Stochastic analysis methods (Fokker-Planck, Langevin, etc.)
%05.10.Gg	Stochastic analysis methods (Fokker-Planck, Langevin, etc.)
%05.20.Gg	Classical ensemble theory
%05.40.-a	Fluctuation phenomena, random processes, noise, and Brownian motion
%Levy flights, 05.40.Fb
%02.50.Ey stochastic processes
%\pacs{05.40.Fb,02.50.Ey}

\maketitle

 \subsection*{A test: Does the superoperator $\mathcal{S}_{\{\mathbf{x}\}}$ have a branch which is conditionally completely positive?}
 
%\underline{Recall}: 
A superoperator $\mathcal{S}$ is \emph{conditionally
completely positive (cc-positive)} \cite{Evans77,Bratteli1984}, if
\begin{equation}
	 \label{eq:ccp}
	\Sigma_\bot \mathcal{S}^\Gamma \Sigma_\bot\geq 0\;,
\end{equation}
where $\Sigma_\bot=\mathbb{1}_{N^2} -\Sigma$ is the projector onto the
orthogonal complement of the maximally
entangled state of the bipartite system ``original system plus ancilla of the same size'', 
\begin{equation}
\ket{\Omega} = \frac{1}{\sqrt{N}}\sum_{j=1}^{N}\ket{jj} \in \mathcal{H}\otimes \mathcal{H},
\label{eq:Bell-omeg}
\end{equation}
 with $\ket{jj}=\ket{j} \otimes \ket{j}$ and $\{\ket{j}\}$ being the canonical basis of the system's Hilbert space $\mathcal{H}$ with dimension $N$. Furthermore, we define $\Sigma = |\Omega\rangle \langle\Omega|$.
$\mathcal{S}^\Gamma$ is the Choi-Jamio\l{}kowski  isomorphism   \cite{Choi75, Jamiolkowski72,BengtssonZyczkowski06,JiangEtAl13},
\begin{equation}
	\mathcal{S}^\Gamma = (\mathcal{S} \otimes \mathbb{1}) \ket{\Omega}\bra{\Omega},
	\label{eq:Choi-def}
\end{equation}
which maps superoperators on $\mathcal{H}$ onto linear operators on the system-plus-ancilla space, $\mathcal{H}\otimes \mathcal{H},$.

%is an involution called ``reshuffling'' (I. Bengtsson and K. \.{Z}yczkowski, \textit{Geometry of Quantum States: An Introduction to Quantum Entanglement} 
%(Cambridge University Press, 2006). \textbf{- Alex: please add this reference}),
%\begin{equation}	
%	\langle i,j|\mathcal{S}^\Gamma|k,l\rangle=\langle i,k|\mathcal{S}|j,l\rangle
%	\langle j,k|\mathcal{S}^\Gamma|l,m\rangle=\langle j,l|\mathcal{S}|k,m\rangle
%\end{equation}	
%It also connects matrix representations 
%of the superoperator $\mathcal{S}$ and its Choi representation (see the main text, p. 4 (first column), and SM section ``Alternative definition of distance to Markovianity''). 

Now we  outline the test, developed and presented in  Refs.~\cite{WolfEtAl08,Cubitt2012}, in a form of a recipe:

\begin{itemize}
\item  Cast  the Floquet map $\mathcal{P}(T)$ into Jordan
normal form,
\begin{align} 
	\mathcal{P}(T) &=    \sum_a \lambda_a \mathcal{M}_{a}  \\ &= \sum_r \lambda_r \mathcal{M}_r + \sum_c \left(
	\lambda_c \mathcal{M}_c + \compcon{\lambda}_c \mathcal{M}_{c*}\right),\label{eq:Jordan}
\end{align}
where $r$ labels the real and $c$ the complex eigenvalues (which come in complex conjugated pairs, $\lambda_c$  and  $\compcon{\lambda}_c$; the number of the pairs is $n_c$), respectively. The
$\mathcal{M}$'s are orthogonal (but typically not self-adjoint) spectral
projectors. 

\item Define a set of operators 
$\mathcal{S}_{\{\mathbf{x}\}} = \frac{1}{T}\log \mathcal{P}(T) = \mathcal{S}_0 + \frac{2\pi i}{T}\sum_{c=1}^{n_c} x_c (\mathcal{M}_c -  \mathcal{{M}}_{c*})$.
Here $\mathcal{S}_0 = \frac{1}{T} \mathrm{Log} \mathcal{P}(T)$ is the principal branch and 
a set of $n_c$ integers $\{\mathbf{x}\} = \{x_1,...,x_{n_c}\} \in \mathbb{Z}^{n_c}$ labels a particular branch of the logarithm.

\item Calculate the operators
\begin{eqnarray}
	V_0&=& \Sigma_\bot \mathcal{S}^\Gamma_0\Sigma_\bot,\\
	V_c&=& \frac{2\pi i}{T}\;
	\Sigma_\bot\big(\mathcal{M}_c-\mathcal{{M}}_{c*}\big)^\Gamma\Sigma_\bot.
\end{eqnarray}
These operators are Hermitian. %(an intermediate check).

\item Define a parametrized set of operators $V_{\{\mathbf{x}\}}  =  V_0 + \sum_{\mathrm{c}} x_c V_c$. 
The task now is to answer the following questions: Is there  a set of $n_c$ integers, $\{\mathbf{x}\} \in \mathbb{Z}^{n_c}$, such that
$V_{\{\mathbf{x}\}} \geq 0$ (i.e., the  spectrum of $V_{\{\mathbf{x}\}}$ contains no negative eigenvalues)? 
If the answer is `yes', the test is passed and the corresponding branch is a Lindbladian.
Otherwise, the effective Floquet Lindbladian does not exist.

At first glance, to test this condition,  we
have to inspect \textit{all} branches, i.e.,\ a countably infinite number of combinations of $n_c$ integers.
Fortunately, the situation is not that hopeless because finding the solution %(or proving its absence) 
for this equation is related to two known programing problems \cite{Ramana1995,Khachiyan2000}. 
When $\{\mathbf{x}\}$ ranges over $\mathbb{R}^{n_c}$, the condition $V_{{\{\mathbf{x}\}}}=0$ out-shapes either zero or a finite volume in which $V_{{\{\mathbf{x}\}}}$ is positive semidefinite. 
To check whether the spectrahedron contains an integer point is a problem of
polynomial complexity with respect to $\mathrm{max}\{|x^0_1|,...,|x^0_{n_c}| \}$, where $\{\mathbf{x}^0 \} \in \mathbb{R}^{n_c}$ is the solution set of $V_{{\{\mathbf{x}\}}}=0$.

When $N=2$ (the case considered in the main text), the problem becomes simple. 
Since there is at most one pair of complex conjugated eigenvalues, the family of operators $V_{x}$ is parametrized
by a single real variable $x$. The minimal eigenvalue of $V_{x}$ is a convex (or concave) real function of a single variable 
and the only task remaining is to check whether there is at least one integer point on the real axis
where the function $V_{x}$  takes a non-negative value.

\end{itemize}

\subsection{Extracting Hamiltonian and jump operators from a Lindblad generator}
In the region where the Floquet Lindbladian $\mathcal{L}_F$ exists, we can extract
its components, that are a time-independent effective Hamiltonian $H_{F}$ and a set of effective jump operators 
$\lbrace A_i \rbrace$.
%, from the underlying map~$\Phi(T)$. 
To this end, we use the fact that  any Lindbladian may be represented in the form \cite{Lindblad1976}
\be
	\mathcal{L}(\varrho) = \varphi(\varrho) - \kappa \varrho -  \varrho \kappa^\dagger,
\ee
where $\kappa \in \mathbb{C}^{N\times N}$ and $\varphi$ is a CP map 
with $\varphi^*(\mathrm{\mathbb{1}}) = \kappa +  \kappa^\dagger$ ($\varphi^*$ denotes the dual map of $\varphi$ 
defined by $\mathrm{tr}[\varphi^*(A)B] = \mathrm{tr}[A\varphi(B)]$ for all linear operators $A, B$).
Then, $i H_F$ is given by the antihermitian part of $\kappa$, $i H_F = \frac{1}{2} (\kappa - \kappa^\dagger)$,
and the Lindblad operators are the Kraus operators of $\varphi$, $\varphi(\cdot) = \sum_{i} A_i \cdot A_i^\dagger$.

Now we may use that in the Basis $B=(\ket{\Omega}, \dots)$,
where $\ket{\Omega}$ % \in \mathbb{C}^n \otimes \mathbb{C}^n$ 
is the maximally entangled state in Eq.~\eqref{eq:Bell-omeg}, the
Choi matrix of $\mathcal{L}$ has the structure
\be
	\begin{split}
	&\mathcal{L}^\Gamma = (\mathcal{L} \otimes \mathbb{1})\left(\ket{\Omega}\bra{\Omega}\right) \\
	&=  \underbrace{\left(\begin{array}{ccccc}
	 	0 &0 & 0 & 0 & \dots\\
		0 & \bullet  & \bullet  & \bullet  &\\
		0 & \bullet &  \bullet &  \bullet &\\
		0 & \bullet  & \bullet  & \bullet  & \\
		\dots &  &   &   &
	  \end{array}\right)}_{\varphi^\Gamma} +\underbrace{
	  \left(\begin{array}{ccccc}
	 	a & \compcon{b} & \compcon{c} & \compcon{d} & \dots\\
		b & 0 & 0  & 0 & \\
		c & 0 & 0  &  0 &\\
		d & 0& 0 & 0 & \\
		\dots &  &   &   &
	  \end{array}\right) }_{-(\kappa \cdot + \cdot \kappa^\dagger)^\Gamma}.
	  \end{split}
\ee
This allows for the direct identification of $\kappa$ and thus the Hamiltonian (up to a global shift of the energies).

For a two-level system, the above representation can be rewritten in the Bell basis
$B=(\ket{\Omega}, \ket{\Sigma}, \ket{\Gamma},  \ket{\Lambda})$, with 
$\ket{\Omega/\Sigma} = \frac{1}{\sqrt{2}} \left( \ket{00} \pm \ket{11} \right)$ and
$\ket{\Gamma/\Lambda} = \frac{1}{\sqrt{2}} \left( \ket{01} \pm \ket{10} \right)$.
This yields
\be
	H_F = \frac{1}{2} \left(\begin{array}{cc}
	 	-\mathrm{Im}( b)& -\mathrm{Im}(c)+i \, \mathrm{Re}(d)\\
		 -\mathrm{Im}(c)-i \, \mathrm{Re}(d) & \mathrm{Im}( b)
	  \end{array}\right).
\ee
%For the dissipative part however, we first have to diagonalize the Choi matrix $\varphi^\Gamma$.
Since $\mathcal{L}_F$ exists, we know that $\varphi^\Gamma$ is positive semi-definite and we can
bring it to the form $\varphi^\Gamma= \sum_i v_i v_i^\dagger$ with vectors 
$v_i = (0, v_{i2}, v_{i3}, \dots)$. Note that these vectors are already the representation of the Lindblad operators $A_i$
 in the sense that $\ket{v_i} = (A_i \otimes \mathbb{1}) \ket{\Omega}$.
 For a two-level system we therefore find 
\be
	A_i = \frac{1}{\sqrt{2}} \left(\begin{array}{cc}
	 	v_{i2}& v_{i3}+v_{i4}\\
		 v_{i3}-v_{i4} & -v_{i2}
	  \end{array}\right).
 \ee

  \subsection*{Definitions of distance to Markovianity}
 
 %\begin{figure}
%	\includegraphics[angle=0,width=0.8\columnwidth]{figS1}
%	\caption{Ratio of the distance to markovianity $d$ that was used in the main text
%	and a similar distance measure $d_\mathrm{P}$ that is also used in the literature \cite{RivasEtAl10}.
%	Parameters correspond to Fig.~\ref{fig:main}(a) of the main text. In the plot, we set ill-defined ratios like $0/0$ to zero.
%	Note that in the corner around $(E,\omega)=(0.1,5)$ the generator is non-markovian so the distance should be finite.
%	However the distance measure $d_\mathrm{P}$ is hard to obtain because the distance values are too small.
%	}
%	\label{fig:comp-meas}
%\end{figure}

One possibility to quantify the  distance to Markovianity is
by diluting the evolution with noise -- until it becomes Markovian \cite{WolfEtAl08}. This means that we add to $\mathcal{S}_{\{\mathbf{x}\}}$ the generator 
$\mathcal{L}_{\chi}= -\chi\Sigma_{\bot}$ (giving rise to the depolarizing channel $\mathcal{P}_{\chi} = \mathrm{e}^{-\chi T} + \frac{\mathbb{1}}{N}[1 - \mathrm{e}^{-\chi T}]$, which when acting alone guides the system into the maximally mixed state in the long-time limit).  
The minimal (over all branches) amount of noise needed to make the evolution Markovian, 
\begin{align}
\mu = \min_{\{\mathbf{x}\}} \min \left\{\chi\geq 0 \vert \mathcal{S}_{\{\mathbf{x}\}} + \mathcal{L}_{\chi} \text{ is cc-positive}  \right\},
\end{align}
defines a `distance to Markovianity' $\mu$. It can also be expressed as  the amount of noise to make $V_{\{\mathbf{x}\}}$ non-negative,  i.e.~\cite{WolfEtAl08}
\begin{align}
\mu = \min_{\{\mathbf{x}\}} \min\left\{\chi\geq 0 \vert V_{\{\mathbf{x}\}} + \frac{\chi}{N} \mathbb{1} \geq 0  \right\}.
\end{align}

An alternative measure to quantify the distance  to Markovianity is
introduced in Ref.~\cite{RivasEtAl10}.
 It is based on the fact that  a given map $\mathcal{P}$ 
 is completely positive iff its Choi representation, Eq.~\eqref{eq:Choi-def}, is positive, $\mathcal{P}^\Gamma  \geq 0$.
 Together with the fact that the map is trace-preserving, $\mathrm{tr}{\mathcal{P}^\Gamma}=1$, one finds that 
 $\vert \vert \mathcal{P}^\Gamma \vert \vert_1 = 1$ iff $\mathcal{P}$ is Markovian  and $\vert \vert \mathcal{P}^\Gamma \vert \vert_1 > 1$
 if it is not (here $ \vert \vert \varrho \vert \vert_1= \mathrm{tr} \sqrt{\varrho^\dagger \varrho}$ is the trace norm). On the level of the generator $\mathcal{L}$,
 $\mathcal{P}(t) = \exp(\mathcal{L}t)$, the derivative of this norm $\vert \vert \mathcal{P}(t)^\Gamma \vert \vert_1$ at $t=0$ can
 be used to define a distance measure
 \begin{equation}
	d_\mathrm{RHP} = \lim_{\varepsilon\to 0}\frac{\vert \vert (\mathbb{1}+\varepsilon \mathcal{L})^\Gamma \vert \vert_1-1}{\varepsilon}.
\end{equation}
Surprisingly, according to our numerics, this measure is identical up to a factor of one half to the measure 
$\mu$. 
For small distances $d_\mathrm{RHP} < 10^{-7}$ the distance measure $d_\mathrm{RHP}$
is hard to obtain numerically, therefore the measure $\mu$ is better in this respect.

 \subsection*{Characteristic decay function of exponential kernel}
 
 In the main text, we show that with a special choice of the
 spectral decomposition of the Kernel superoperator $\mathcal{K}$ 
 the problem of engineering an effective evolution with a time-homogeneous memory
 kernel can be reduced to solving a scalar integro-differential equation for
 the characteristic decay functions $h_{a}(t)$. Here we solve this equation.
 
Using the ansatz $\mathcal{K}= \sum_a \eta_a \mathcal{M}_{a}$ 
we find an evolution operator of the form 
$\tilde{\mathcal{P}}(t) = \sum_a h_{a}(t) \mathcal{M}_{a}$. Plugging this into the equation of motion for $\tilde{\mathcal{P}}(t)$
we find a scalar equation
\begin{align}\label{eq:kernel-scalar}
	\partial_t h_{a}(t) = \Gamma \int_{0}^t \mathrm d \tau \, \mathrm{e}^{\Gamma (\tau-t)} \eta_a h_{a}(\tau)
\end{align}
% in the main text 
(we set $\Gamma =1/\tau_\text{mem}$ for convenience). It can be transformed into a second order differential equation, by taking its derivative,
\begin{align}
	\begin{split}
	\partial^2_t h_{a}(t) &= -\Gamma^2  \int_{0}^t \mathrm d \tau \,  \mathrm{e}^{-\Gamma(t-\tau)}\, \eta_a h_{a}(\tau) \\
	&\qquad + \Gamma \mathrm{e}^{-\Gamma t} \, \mathrm{e}^{+\Gamma t}  \eta_a h_{a}(t)
	 \end{split}
	 \\
	&= - \Gamma  \partial_t h_{a}(t) +\Gamma  \eta_a h_{a}(t),
\end{align}
where additionally we have to satisfy the boundary conditions $h_a(0) = 1, \ \partial h_a(t)  \vert_{t=0}= 0$ 
[by setting $t=0$ in Eq.~\eqref{eq:kernel-scalar}].

This linear homogeneous second order  differential equation can be solved by exponential ansatz
\begin{equation}
	h_{a}(t) = \mathrm{e}^{\mu_a t}
\end{equation}
which leads to the characteristic polynomial
\begin{equation}
	(\mu_a^2 + \Gamma \mu_a - \Gamma \eta_a ) h_{a}(t) = 0
\end{equation}
which is solved by
\begin{equation}
	\mu_a^{\pm} = -\Gamma/2 \pm \Gamma_a
\end{equation}
with the complex root $\Gamma_a = \sqrt{\Gamma^2/4 + \Gamma \eta_a}$. So the general solution takes the form
\begin{equation}
	h_{a}(t) = \mathrm{e}^{-\Gamma t/2} \left( \alpha \mathrm{e}^{\Gamma_a t} + \beta \mathrm{e}^{-\Gamma_a t} \right).
\end{equation}
By imposing $h_a(0) = 1, \, \partial h_a(t)  \vert_{t=0}= 0$ we find
\begin{equation}
	\alpha+\beta=1, \quad \text{and} \quad -\frac{\Gamma}{2} (\alpha+\beta) +  \Gamma_a (\alpha-\beta)= 0.
\end{equation}
This is solved by
\begin{equation}
	2\alpha = 1 + \frac{\Gamma}{2\Gamma_a}
\end{equation}
so that we finally get
\begin{equation}
	h_{a}(t) = \mathrm{e}^{-\Gamma t/2} \left[ \frac{1}{2}\left(  \mathrm{e}^{\Gamma_a t} + \mathrm{e}^{-\Gamma_a t} \right) + \frac{\Gamma}{4\Gamma_a}\left(  \mathrm{e}^{\Gamma_a t} - \mathrm{e}^{-\Gamma_a t} \right)  \right].
	\label{eq:h_a-explicit}
\end{equation}

\begin{figure}\includegraphics[angle=0,width=0.65\columnwidth]{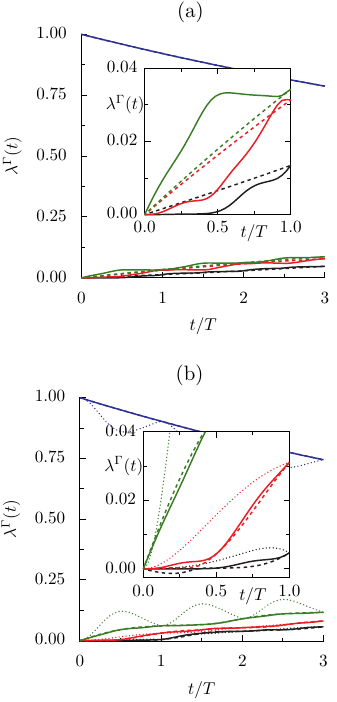}
	\caption{Instantaneous eigenvalues $\lambda^\Gamma(t)$ of the Choi matrix of
	the dynamical map $\mathcal{P}(t)$ (solid lines) and of the effective semigroup $\exp(t \mathcal{L}^{c})$ (dashed lines)
	for the two-level model with $\gamma=0.01, \varphi=0$ and (a) $ \omega=1.5, E=1.5$ as well as (b) $ \omega=1.2, E=0.75$.
	$\mathcal{L}^{c}=\mathrm{log}(\mathcal{P}(T))/T$ is chosen from the branch that is closest to Markovianity.
	By construction, both evolutions coincide at integer multiples of the period, $t = nT$.
	The inset shows a zoom into the three smallest eigenvalues  and the first period.
	The evolution is CPTP only if all eigenvalues of the Choi matrix are non-negative for all times.
	By construction $\mathcal{P}(t)$ is CPTP. The semigroup evolution in (a) is CPTP, thus $\mathcal{L}_F$
	exists, but in (b) it is not CPTP, thus no $\mathcal{L}_F$ exists.
	The dotted lines stem from the evolution with the designed exponential kernel. Even though there is no time-local effective
	evolution with a  time-independent generator $\mathcal{L}_F$, the time-homogeneous time-non-local evolution with the designed kernel
	is CPTP at all times and coincides with $\mathcal{P}$ at the full period $T$ (dotted lines).
	}
	\label{fig:Eval-Choi}
\end{figure}

 \subsection*{Numerically stable procedure to find eigenvalues of  kernel superoperator $\mathcal{K}$}
 
% Solving the nonlinear equation $h_{a}(T) = \lambda_a$ for $\lambda^K_a$ can in general not be performed analytically.
In order to have the stroboscopic identity 
\begin{align}
	\tilde{\mathcal{P}}(T) = {\mathcal{P}}(T)
\end{align}
we need to solve the equation $h_{a}(T) = \lambda_a$  for $\eta_a$.
 However, in general, this nonlinear equation % $h_{a}(T) = \lambda_a$ 
 cannot be solved analytically.
Nevertheless, for the steady state subspace, $\lambda_a=1$, we directly infer that $\eta_a = 0$ is a solution. 
%Note that one eigenvalue $0$ is required, 
Since $\mathcal{K}=\mathcal{K}{(\Gamma)}$ has to preserve hermiticity %be a valid generator 
%and 
it is of the form
\begin{equation}
	\mathcal{K}[\cdot] = 0 \cdot \mathcal{M}_{SS}[\cdot] + \sum_r \eta_r \mathcal{M}_{r}[\cdot] + \sum_c (\eta_c \mathcal{M}_{c}[\cdot] + \compcon{\eta}_c \mathcal{{M}}_{{c*}}[\cdot])
\end{equation}
with real eigenvalues $\eta_r$ and pairs of complex conjugated eigenvalues $\eta_c, \compcon{\eta}_c$.
One way  to determine the remaining $\eta_a$ would be
to use a numerical root finding algorithm. However, the stroboscopic identity $h_{a}(T) = \lambda_a$  has generally infinitely many
solutions in the complex plane, so that we observe that a root finding algorithm can converge into solutions with large imaginary part
(which generally yields a $\mathcal{K}$ that is not a valid generator; similar to the Markovian case we suspect that
branches~$x$ that are far off from the principle branch do not give a valid generator anymore).
 %e.g.~we are never really sure that although a real solution $\lambda_r$ 
%exists we maybe ran into one of the complex ones.

A numerical way around this problem is expanding the equation $h_{a}(T) = \lambda_a$  in a power series, then cutting it off at some
index, so we have a polynomial equation, where all roots of the polynomial can be evaluated from the numerics. Rewriting
$h_{a}(T) = \lambda_a$ in a power series then gives
\begin{equation}
	\sum_{n=0}^{\infty} \left( \frac{(\Gamma_a T)^{2n}}{(2n)!} + \frac{\Gamma}{4 \Gamma_a} \frac{(\Gamma_a T)^{2n+1}}{(2n+1)!} \right) = \lambda_a \mathrm{e}^{\Gamma T/2}.
\end{equation}
Using the definition of $\Gamma_a$ we find
\begin{equation}
	\sum_{n=0}^{\infty} (\Gamma^2/4 +  \Gamma  \eta_a)^{n} \, T^{2n} \left( \frac{1}{(2n)!} + \frac{\Gamma}{4} \frac{T}{(2n+1)!} \right) =\lambda_a \mathrm{e}^{\Gamma T/2},
\end{equation}
where we cut off the power series at some index $n_0$, solve for all solutions $z=\Gamma^2/4 +  \Gamma  \eta_a$, and then
regain all possible $\eta_a = (z-\Gamma^2/4)/\Gamma$. In this way we %in principle,
 again find infinitely many candidates for $\mathcal{K}$.
For the case where $\lambda_a$ is real we only may take the one root $z$ where $\eta_a$ is real, but for the complex pair 
$\eta_c$ there is no restriction (apart from the requirement of keeping complex conjugated pairs), so we can choose any complex solution $\eta_c$. 
However only for solutions with a small absolute value, we may cut off the sum at index $n_0$. In order to avoid inaccuracies, we thus
restrict ourselves to a few solutions $\eta_c$  with small imaginary part. Still, for the two-level system 
one can always find a  parameter $\Gamma$ such that a valid kernel evolution exists (one still has to guarantee CPTP of  $\tilde{\mathcal{P}}(t)$ at all times $t\in [0, T]$ as described in the main text)  and we observe that in most cases it
suffices to consider the solution $\eta_c$ with the smallest imaginary part.

In Fig.~\ref{fig:Eval-Choi} we show two examples for the driven-dissipative two-level system. It is instructive to analyze 
the eigenvalues $\lambda^\Gamma$ of the Choi image  of the evolution operator $\mathcal{P}(t)$. The evolution is CPTP only if all these  eigenvalues are
non-negative. The parameters in Fig.~\ref{fig:Eval-Choi}(a) lie in the region where a Floquet Lindbladian exists, therefore there is a
semigroup evolution (dashed lines) that coincides with $\mathcal{P}(nT)$, $n \in \mathbb{N}_0$, and is CPTP for all times. For
the parameters in Fig.~\ref{fig:Eval-Choi}(b) such a semigroup evolution does not exist. We show the semigroup 
evolution that is closest to a CPTP evolution (in the sense of the distance measure $\mu$).
However, there exists an evolution with a time-homogeneous exponential kernel in the sense of Eq.~\eqkernel in the main text  (dotted lines) [using
that we erase the memory at stroboscopic times with the Heaviside function], that we found with the procedure above,
which coincides with $\mathcal{P}(nT)$, $n \in \mathbb{N}_0$, and is CPTP for all times.

 \subsection*{Remarks on guaranteeing CPTP for the  non-Markovian evolution}

\begin{figure}{\centering \includegraphics[angle=0,width=1.1\columnwidth]{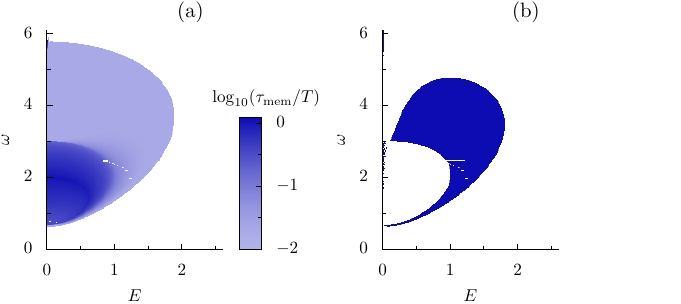}}
	\caption{(a)~Shortest memory time  $\mathcal{\tau}_\mathrm{mem}$ as function of $E$ and $\omega$. 
	Different to Fig.~3 in the main text, here we %in analogy to Fig.~3 of the main text,
	only restrict $\mathcal{K}$ to Lindblad form.
	%require solely that $\mathcal{K}$ is a valid Lindblad operator. 
	The resulting values of $\mathcal{\tau}_\mathrm{mem}$ look qualitatively very similar to
	the exact values in Fig.~3 of the main text. (There are some isolated white points where the numerics did not converge.)
	(b)~Regions in which the corresponding evolution with this Lindbladian $\mathcal{K}$ as 
	kernel superoperator	 is a CPTP evolution (white), or is no CPTP evolution (blue).
	}
	\label{fig:CPTP-from-Lindblad}
\end{figure}

Note that for the non-Markovian evolution, Eq.~\eqkernel in the main text, we guarantee CPTP not by imposing conditions on 
the generator $\mathcal{K}$, but on the evolution $\tilde{\mathcal{P}}(t)$ itself. In the Markovian limit, $\tau_\mathrm{mem}\to 0$,
such a condition exists: The generator has to obey Lindblad form. In the non-Markovian regime such a simple criterion 
is not known. A naive idea would be that for short, but finite $\tau_\mathrm{mem}$, this criterion carries over, and by guaranteeing 
that $\mathcal{K}$ is a valid Lindbladian we find an evolution that is `almost' CPTP.

Motivated by this idea, in Fig.~\ref{fig:CPTP-from-Lindblad}(a) we show the shortest memory time $\tau_\mathrm{mem}$
for which $\mathcal{K}$ is a valid Lindbladian by performing the test for condition (ii). 
We find a good qualitative agreement to the exact values in  Fig.~3 of the main text. 
This means that choosing $\mathcal{K}$ to be of Lindblad form is a good and easy test to determine at least the order 
of magnitude of the required memory $\tau_\mathrm{mem}$.
In Fig.~\ref{fig:CPTP-from-Lindblad}(b) we plot whether this test is also quantitatively correct, i.e.~whether
the  non-Markovian evolution, Eq.~\eqkernel\!\!, with this Lindbladian kernel $\mathcal{K}$ is giving rise to an evolution $\tilde{\mathcal{P}}(t)$ that is CPTP (white region) or not (blue region).
%However, for our purposes 
The following observation  is noteworthy: Although one naively expects that guaranteeing that  $\mathcal{K}$ is a valid Lindbladian 
is a good test for short memory times $\tau_\mathrm{mem}$ only, we find the contrary, namely that this test is a good test
especially for long memory times $\tau_\mathrm{mem}$.

Why this is the case can be anticipated by the following fact: As we observe in Fig.~\ref{fig:Eval-Choi}(a), complete positivity
is often violated for short times $t$ only.
For $t \ll \tau_\mathrm{mem}$ we may approximate
\begin{equation}
\mathrm{e}^{\pm \Gamma_a t} \approx 1\pm \Gamma_a t+ \frac{1}{2} \Gamma_a^2 t^2
\end{equation}
as well as
\begin{equation}
\mathrm{e}^{-\Gamma t/2} \approx 1- \frac{1}{2} \Gamma {t}+ \frac{1}{8} \Gamma^2 t^2.
\end{equation}
Plugging this into Eq.~\eqref{eq:h_a-explicit}, we find
\begin{align}
h_a(t) &\approx 1 + 0  t + \left( \frac{\Gamma^2}{8} - \frac{\Gamma^2}{4} +  \frac{\Gamma_a^2}{2}\right) t^2 = 1+\frac{\Gamma \eta_a}{2} t^2\\
&\approx \exp\left(\frac{\Gamma \eta_a}{2} t^2\right).
\end{align}
As a result, for short times $t \ll \tau_\mathrm{mem}$ we may rewrite the evolution as
\begin{equation}
	\tilde{\mathcal{P}}(t)[\cdot]  = \sum_a h_a(t) \mathcal{M}_{a}[\cdot] \approx \exp\left(\frac{\mathcal{K}}{2\tau_\mathrm{mem}} t^2\right)[\cdot].
	\label{eq:nonmarkov-eff-semigoup}
\end{equation}
Thus, for $t \ll \tau_\mathrm{mem}$ the evolution is effectively given by a quantum  semigroup evolution, 
but with rescaled time ${t}'=t^2/2\tau_\mathrm{mem}$.
As a result, we can guarantee a CPTP evolution for short times %by guaranteeing that $\mathcal{K}$ is a valid Lindblad generator.
 if $\mathcal{K}$ is of the Lindblad form.
For large values of $ \tau_\mathrm{mem}$ the approximation in Eq.~\eqref{eq:nonmarkov-eff-semigoup} holds for a
substantial fraction of the period $T$, so that guaranteeing that Eq.~\eqref{eq:nonmarkov-eff-semigoup} is CPTP seems  often to be enough to guarantee CPTP for the full period.

%From some numerical examples that we studied we see that already the value $\lambda_a$ with the smallest imaginary part can lead to 
%a valid generator for the kernel evolution, even in regimes where there is no Floquet Lindbladian $\mathcal{L}_F$.
 \bibliography{mybib}
 
% \begin{thebibliography}{2} 
 
%\bibitem{Lindblad1976}  G. Lindblad,Comm. Math. Phys. \textbf{48}, 119 (1976).

%\bibitem{RivasEtAl10} \'{A}. Rivas, S. F. Huelga, and M. B. Plenio, Phys. Rev. Lett. \textbf{105}, 050403 (2010).
%\bibitem{Choi75} M.-D. Choi, Linear Algebr. Appl.  \textbf{10}, 285 (1975).
%\bibitem{Jamiolkowski72} A. Jamiokowski, Rep. Math. Phys. \textbf{3}, 275 (1972)
%\bibitem{wolf1} M. M. Wolf, J. Eisert, T. S. Cubitt, and J. I. Cirac, Phys. Rev. Lett. \textbf{101}, 150402 (2008).

%\end{thebibliography}